\documentclass[prl,preprint,amsmath,amssymb]{revtex4}
\usepackage{graphicx}
\usepackage[justification=centering]{caption}

\begin{document}

\title{Improved monogamy relations with concurrence of assistance and negativity of assistance
for multiqubit W-class states}

\author{Zhi-Xiang Jin$^{1}$}
\author{Shao-Ming Fei$^{1,2}$}
\author{Xianqing Li-Jost$^{2,3}$}
\affiliation{$^1$School of Mathematical Sciences, Capital Normal University,
Beijing 100048, China\\
$^2$Max-Planck-Institute for Mathematics in the Sciences, 04103 Leipzig, Germany\\
$^3$School of Mathematics and Statistics, Hainan Normal University, Haikou 571158, China}
\bigskip

\begin{abstract}

Monogamy relations are important for characterizing the distributions of entanglement in multipartite systems. We investigate the monogamy relations satisfied by the concurrence of assistance and the negativity of assistance for multiqubit generalized $W$-class states. Analytical monogamy inequalities are presented for both concurrence of assistance and negativity of assistance, which are shown to be tighter than the existing ones. Detailed examples have been presented.

\end{abstract}

\maketitle

\section{INTRODUCTION}
Quantum entanglement \cite{MAN,RPMK,FMA,KSS,HPB,HPBO,JIV,CYSG} is one of the most important features of quantum mechanics. It has the essential  differences as for the quantum entanglement and classical correlations. Among them, monogamy inequality piays a important role, which characterizes a subsystem entangled with one of the subsystems limits its entanglement with the remaining subsystems. 
The monogamy inequality implies the relations, such as entanglement, the distribution of quantum correlation in multipartite systems. Monogamy property also do much help for security in quantum key distribution \cite{MP}.
  									
For a tripartite system $A$, $B$ and $C$, the authors \cite{ckw61052306} show that there is a trade-off between $A's$ entanglement with $B$ and its entanglement with $C$. In \cite{kw69022309}, the authors present a simple identity which captures the trade-off between entanglement and classical correlation, which can be used to derive rigorous monogamy relations. They also proved various monogamous trade-off relations for other entanglement measures and correlation measures. In \cite{tf96.220503}, the author proved the longstanding conjecture of Coffman, Kundu, and Wootters \cite{ckw61052306} that the distribution of bipartite quantum entanglement, measured by the tangle $\tau$, amongst $n$ qubits satisfies a tight inequality: $\tau{\rho_{A_1A_2}}+\tau{\rho_{A_1A_3}}+\cdots+\tau{\rho_{A_1A_n}}\leq \tau{\rho_{A_1|A_2\cdots A_n}}$,
where $\tau{\rho_{A_1|A_2\cdots A_n}}$ denotes the bipartite quantum entanglement measured by the tangle under the bipartition $A_1$ and $A_2 A_3\cdots A_n$.

Recently, some monogamy-type inequalities have been investigated, such as tangle and the squared concurrence  \cite{KSJ,KSJB}. In Ref. \cite{ZF}, the general monogamy inequalities of the $\alpha$-th power of concurrence and entanglement of formation are presented for $N$-qubit states. However, the concurrence of assistance does not satisfy monogamy relations for general quantum states. Therefore, special classes of quantum states have been taken into account for monogamy relations satisfied by the concurrence of assistance. The monogamy relations for the $x$-power of concurrence of assistance for the generalized multiqubit $W$-class states have been derived in \cite{ZXN}. In \cite{jzx1}, a tighter monogamy relation of quantum entnglement for multiqubit $W$-class states has been presented.

In this paper, we obtain tighter monogamy relations based on concurrence of assistance. We present the $x$-th $(x\geq2)$ power of the concurrence of assistance satisfy the monogamy inequality, which are tighter than those in \cite{ZXN,jzx1}.
Moreover, general monogamy inequalities for the $x$-power of negativity of assistance are given for the generalized multiqubit $W$-class states, which are also better than that in \cite{ZXN,jzx1}.

\section{improved MONOGAMY RELATIONS FOR CONCURRENCE OF ASSISTANCE}

For a pure state $|\psi\rangle_{AB}$ in vector space $H_A\otimes H_B$, the concurrence is defined by $C(|\psi\rangle_{AB})=\sqrt{{2\left[1-\mathrm{Tr}(\rho_A^2)\right]}}$ \cite{AU,PR,SA},
where $\rho_A$ is the reduced density matrix, $\rho_A=\mathrm{Tr}_B(|\psi\rangle_{AB}\langle\psi|)$. The concurrence for a bipartite mixed state $\rho_{AB}$ is defined by the convex roof extension
$C(\rho_{AB})=\min_{\{p_i,|\psi_i\rangle\}}\sum_ip_iC(|\psi_i\rangle)$,
where the minimum is taken over all possible pure state decompositions of $\rho_{AB}=\sum_ip_i|\psi_i\rangle\langle\psi_i|$, with $p_i\geq0$ and $\sum_ip_i=1$ and $|\psi_i\rangle\in H_A\otimes H_B$.

For an $N$-qubit state $\rho_{AB_1\cdots B_{N-1}}\in {H}_A\otimes {H}_{B_1}\otimes\cdots\otimes {H}_{B_{N-1}}$, the concurrence $C(\rho_{A|B_1\cdots B_{N-1}})$ of the state $\rho_{A|B_1\cdots B_{N-1}}$, viewed as a bipartite state under the bipartition $A$ and $B_1,B_2,\cdots, B_{N-1}$, satisfies \cite{ZF}
\begin{eqnarray*}\label{mo1}
C^x(\rho_{A|B_1,B_2\cdots,B_{N-1}})\geq
C^x(\rho_{AB_1})+C^x(\rho_{AB_2})+\cdots+C^x(\rho_{AB_{N-1}}),
\end{eqnarray*}
for $x\geq2$, where $\rho_{AB_i}=\mathrm{Tr}_{B_1\cdots B_{i-1}B_{i+1}\cdots B_{N-1}}(\rho_{AB_1\cdots B_{N-1}})$. The above monogamy relation is improved such that for $x\geq2$, if $C(\rho_{AB_i})\geq C(\rho_{A|B_{i+1}\cdots B_{N-1}})$ for $i=1, 2, \cdots, m$, and
$C(\rho_{AB_j})\leq C(\rho_{A|B_{j+1}\cdots B_{N-1}})$ for $j=m+1,\cdots,N-2$,
$\forall$ $1\leq m\leq N-3$, $N\geq 4$, then \cite{JZX},
\begin{eqnarray}\label{l1}
&&C^x(\rho_{A|B_1B_2\cdots B_{N-1}})\geq \nonumber\\
&&C^x(\rho_{AB_1})+\frac{x}{2} C^x(\rho_{AB_2})+\cdots+\left(\frac{x}{2}\right)^{m-1}C^x(\rho_{AB_m})\nonumber\\
&&+\left(\frac{x}{2}\right)^{m+1}\left(C^x(\rho_{AB_{m+1}})+\cdots+C^x(\rho_{AB_{N-2}})\right)\nonumber\\
&&+\left(\frac{x}{2}\right)^{m}C^x(\rho_{AB_{N-1}}).
\end{eqnarray}
In \cite{jll}, (\ref{l1}) is further improved such that for $x\geq2$, one has
\begin{eqnarray}\label{jzx1}
&&C^x(\rho_{A|B_1B_2\cdots B_{N-1}})\geq \nonumber\\
&&C^x(\rho_{AB_1})
 +h C^x(\rho_{AB_2})+\cdots+h^{m-1}C^x(\rho_{AB_m}) \nonumber\\
 &&+h^{m+1}\left(C^x(\rho_{AB_{m+1}})
 +\cdots+C^x(\rho_{AB_{N-2}})\right)\nonumber\\
&&+h^{m}C^x(\rho_{AB_{N-1}})
\end{eqnarray}
for all $x\geq2$, where $h=2^\frac{x}{2}-1$.

For a tripartite pure state $|\psi\rangle_{ABC}$, the concurrence of assistance
is defined by \cite{TFS, YCS}
\begin{eqnarray*}
C_a(|\psi\rangle_{ABC})\equiv C_a(\rho_{AB})=\mathrm{max}_{\{p_i,|\psi_i\rangle\}}\sum_ip_iC(|\psi_i\rangle),
\end{eqnarray*}
where the maximum is taken over all possible decompositions of $\rho_{AB}=\mathrm{Tr}_C(|\psi\rangle_{ABC}\langle\psi|)=\sum_ip_i|\psi_i\rangle_{AB}\langle\psi_i|.$ When $\rho_{AB}$ is a pure state, then one has $C(|\psi\rangle_{AB})=C_a(\rho_{AB})$.

Different from the Coffman-Kundu-Wootters inequality satisfied by the concurrence, the concurrence of assistance does not satisfy the monogamy relations in general. However, for $N$-qubit generalized $W$-class state, $|\psi\rangle_{AB_1\cdots B_{N-1}}\in H_A\otimes H_{B_1}\otimes\cdots\otimes H_{B_{N-1}}$ defined by
\begin{eqnarray}\label{gw}
|\psi\rangle_{AB_1\cdots B_{N-1}}=a|10\cdots0\rangle+b_1|01\cdots0\rangle+\cdots+b_{N-1}|00\cdots1\rangle,
\end{eqnarray}
with $|a|^2+\sum_{i=1}^{N-1}|b_i|^2=1$,
one has \cite{ZXN},
\begin{eqnarray}\label{la2}
C(\rho_{AB_i})=C_a(\rho_{AB_i}),~~~~i=1,2,...,N-1,
\end{eqnarray}
where $\rho_{AB_i}=\mathrm{Tr}_{B_1\cdots B_{i-1}B_{i+1}\cdots B_{N-1}}(|\psi\rangle_{AB_1\cdots B_{N-1}}\langle\psi|)$, and the concurrence of assistance $C_a(|\psi\rangle_{A|B_1\cdots B_{N-1}})$ satisfies
the monogamy inequality \cite{ZXN},
\begin{equation}\label{ZXN1}
  C_a^x(|\psi\rangle_{A|B_1,B_2\cdots,B_{N-1}})\geq C_a^x(\rho_{AB_1})+C_a^x(\rho_{AB_2})+\cdots+C_a^x(\rho_{AB_{N-1}}),
\end{equation}
for $x\geq2$.
(\ref{ZXN1}) has been further improved such that for $x\geq2$, if $C(\rho_{AB_i})\geq C(\rho_{A|B_{i+1}\cdots B_{N-1}})$ for $i=1, 2, \cdots, m$, and
$C(\rho_{AB_j})\leq C(\rho_{A|B_{j+1}\cdots B_{N-1}})$ for $j=m+1,\cdots,N-2$,
$\forall$ $1\leq m\leq N-3$, $N\geq 4$, then \cite{jzx1},
\begin{eqnarray}\label{jzx2}
&&C_a^x(|\psi\rangle_{A|B_1B_2\cdots B_{N-1}})\geq \nonumber\\
&&C_a^x(\rho_{AB_1})+\frac{x}{2} C_a^x(\rho_{AB_2})+\cdots+\left(\frac{x}{2}\right)^{m-1}C_a^x(\rho_{AB_m})\nonumber\\
&&+\left(\frac{x}{2}\right)^{m+1}\left(C_a^x(\rho_{AB_{m+1}})+\cdots+C_a^x(\rho_{AB_{N-2}})\right)\nonumber\\
&&+\left(\frac{x}{2}\right)^{m}C_a^x(\rho_{AB_{N-1}}).
\end{eqnarray}

In fact, as a kind of characterization of the entanglement distribution among the subsystems, the monogamy inequalities satisfied by the concurrence of assistance can be further refined and become tighter.

{[\bf Theorem 1]}.
Let $\rho_{AB_{j_1}\cdots B_{j_{m-1}}}$ denote the $m$-qubit reduced density matrix of the $N$-qubit generalized $W$-class state $|\psi\rangle_{AB_1\cdots B_{N-1}}\in H_A\otimes H_{B_1}\otimes\cdots\otimes H_{B_{N-1}}$. If $C(\rho_{AB_{j_i}})\geq C(\rho_{AB_{j_{i+1}}\cdots B_{j_{m-1}}})$ for $i=1,2,\cdots t$, and $C(\rho_{AB_{j_k}})\leq C(\rho_{AB_{j_{k+1}}\cdots B_{j_{m-1}}})$ for $k=t+1,\cdots,{m-2}$, $\forall$ $1\leq t\leq {m-3},~m\geq 4$, the concurrence of assistance satisfies
\begin{eqnarray}\label{th1}
&&C_a^x(\rho_{A|B_{j_1}\cdots B_{j_{m-1}}})\geq C_a^x(\rho_{AB_{j_1}}) \nonumber\\
&&+h C_a^x(\rho_{AB_{j_2}})+\cdots+h^{t-1}C_a^x(\rho_{AB_{j_t}}) \nonumber\\
&&+h^{t+1}\left(C_a^x(\rho_{AB_{j_{t+1}}})+\cdots+C_a^x(\rho_{AB_{j_{m-2}}})\right)\nonumber\\
&&+h^tC_a^x(\rho_{AB_{j_{m-1}}})
\end{eqnarray}
for all $x\geq2$, where $h=2^\frac{x}{2}-1$.

{\sf [Proof].}
For the $N$-qubit generalized $W$-class states $|\psi\rangle_{AB_1\cdots B_{N-1}}$, according to the definitions of $C(\rho)$ and $C_a(\rho)$, one has $C_a(\rho_{A|B_{j_1}\cdots B_{j_{m-1}}})\geq C(\rho_{A|B_{j_1}\cdots B_{j_{m-1}}})$. When $x\geq2$, we have
\begin{eqnarray}\label{pf1}
C_a^x(\rho_{A|B_{j_1}\cdots B_{j_{m-1}}})
&&\geq C^x(\rho_{A|B_{j_1}\cdots B_{j_{m-1}}}) \geq C^x(\rho_{AB_{j_1}}) \nonumber\\
&&+h C^x(\rho_{AB_{j_2}})+\cdots+h^{t-1}C^x(\rho_{AB_{j_t}}) \nonumber\\
&&+h^{t+1}\left(C^x(\rho_{AB_{j_{t+1}}})+\cdots+C^x(\rho_{AB_{j_{m-2}}})\right)\nonumber\\
&&+h^tC^x(\rho_{AB_{j_{m-1}}})\nonumber\\
&&=C_a^x(\rho_{AB_{j_1}})
+hC_a^x(\rho_{AB_{j_2}})+\cdots+h^{t-1}C_a^x(\rho_{AB_{j_t}}) \nonumber\\
&&+h^{t+1}\left(C_a^x(\rho_{AB_{j_{t+1}}})+\cdots+C_a^x(\rho_{AB_{j_{m-2}}})\right)\nonumber\\
&&+h^tC_a^x(\rho_{AB_{j_{m-1}}}),
\end{eqnarray}
where we have used in the first inequality the relation $a^x\geq b^x$ for $a\geq b\geq 0,~x\geq 2$. The second inequality is due to (\ref{jzx1}). The equality is due to (\ref{la2}). \hfill \rule{1ex}{1ex}

As for $x\geq 2$, $h^t\geq (x/2)^t$ for all $1\leq t\leq m-3$, comparing with
Eqs. (\ref{ZXN1}) and (\ref{jzx2}), our formula (\ref{th1}) in Theorem 1 gives a tighter monogamy relation with larger lower bound. In Theorem 1 we have assumed that
some $C(\rho_{AB_{j_i}})\geq C(\rho_{AB_{j_{i+1}}\cdots B_{j_{m-1}}})$ and some
$C(\rho_{AB_{j_k}})\leq C(\rho_{AB_{j_{k+1}}\cdots B_{j_{m-1}}})$ for the $N$-qubit generalized $W$-class states.
If all $C(\rho_{AB_{j_i}})\geq C(\rho_{AB_{j_{i+1}}\cdots B_{j_{m-1}}})$ for $i=1, 2, \cdots, {m-2}$, then we have
the following conclusion:

{\bf [Theorem 2]}.
If $C(\rho_{AB_{j_i}})\geq C(\rho_{AB_{j_{i+1}}\cdots B_{j_{m-1}}})$ for $i=1, 2, \cdots, {m-2}$, we have
\begin{eqnarray}\label{th2}
C_a^x(\rho_{A|B_{j_1}\cdots B_{j_{m-1}}})\geq C_a^x(\rho_{AB_{j_1}}) +hC_a^x(\rho_{AB_{j_2}})+\cdots+h^{m-2}C_a^x(\rho_{AB_{j_{m-1}}})
\end{eqnarray}
for all $x\geq2$, where $h=2^\frac{x}{2}-1$.

{\it Example 1}. Let us consider the 4-qubit generlized $W$-class state,
\begin{eqnarray}\label{W4}
|W\rangle_{AB_1B_2B_3}=\frac{1}{2}(|1000\rangle+|0100\rangle+|0010\rangle+|0001\rangle).
\end{eqnarray}
We have $C_a^x(|W\rangle_{A|B_1B_2B_3})=(\frac{\sqrt{3}}{2})^x$. From our result (\ref{th1}) we have $C_a^x(|W\rangle_{A|B_1B_2B_3})\geq \left[2\cdot 2^\frac{x}{2}-1\right](\frac{1}{2})^x$, from (\ref{jzx2}) one has $C_a^x(|W\rangle_{A|B_1B_2B_3})\geq (x+1)(\frac{1}{2})^x$, and from (\ref{ZXN1}) one has $C_a^x(|W\rangle_{A|B_1B_2B_3})\geq 3(\frac{1}{2})^x$, $x\geq2$. One can see that our result is better than that in \cite{ZXN} and \cite{jzx1} for $x\geq2$, see Fig. 1.

\begin{figure}
  \centering
  \includegraphics[width=13cm]{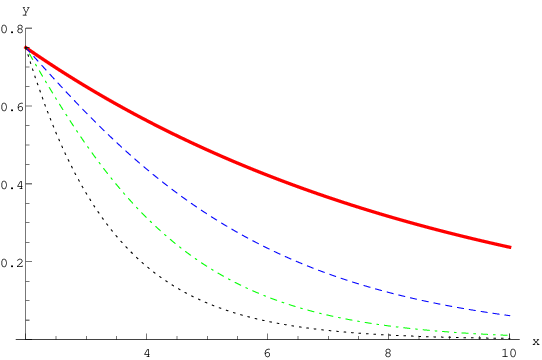}\\
  \caption{$y$ is the value of $C_a(|W\rangle_{A|B_1B_2B_3})$. Solid (red) line is the exact value of $C_a(|W\rangle_{A|B_1B_2B_3})$, dashed (blue) line is the lower bound of $C_a(|W\rangle_{A|B_1B_2B_3})$ in (\ref{th1}), dot dashed (green) line is the lower bound in \cite{jzx1}, and dotted (black) line is the lower bound in \cite{ZXN} for $x\geq2$.}\label{1}
\end{figure}

\section{IMPROVED MONOGAMY RELATIONS FOR NEGATIVITY OF ASSISTANCE}
For a bipartite state $\rho_{AB}$ in $H_A\otimes H_B$, the negativity is defined by \cite{GRF},
$N(\rho_{AB})=(||\rho_{AB}^{T_A}||-1)/2$,
where $\rho_{AB}^{T_A}$ is the partially transposed $\rho_{AB}$ for the subsystem $A$, $||X||$ denotes the trace norm of $X$, i.e $||X||=\mathrm{Tr}\sqrt{XX^\dag}$.
Negativity is a computable measure of entanglement, and is a convex function of $\rho_{AB}$. It vanishes iff $\rho_{AB}$ is separable for the $2\otimes2$ and $2\otimes3$ systems \cite{MPR}. We use the following definition of negativity for convenience, $ N(\rho_{AB})=||\rho_{AB}^{T_A}||-1$.
For a pure state $|\psi\rangle_{AB}$, the negativity $ N(\rho_{AB})$ is defined by
$N(|\psi\rangle_{AB})=2\sum_{i<j}\sqrt{\lambda_i\lambda_j}=(\mathrm{Tr}\sqrt{\rho_A})^2-1$,
where $\lambda_i$ are the eigenvalues for the reduced density matrix $\rho_A$ of $|\psi\rangle_{AB}$. For a mixed state $\rho_{AB}$, the convex-roof extended negativity (CREN) is defined by
\begin{equation}\label{nc}
 N_c(\rho_{AB})=\mathrm{min}\sum_ip_iN(|\psi_i\rangle_{AB}),
\end{equation}
where the minimum is taken over all possible pure state decompositions $\{p_i,~|\psi_i\rangle_{AB}\}$ of $\rho_{AB}$. CREN presents a perfect discrimination of  bound entangled states and separable states in bipartite systems \cite{PH,WJM}. For a mixed state $\rho_{AB}$, the
convex-roof extended negativity of assistance (CRENOA) is defined as \cite{JAB}
\begin{equation}\label{na}
 N_a(\rho_{AB})=\mathrm{max}\sum_ip_iN(|\psi_i\rangle_{AB}),
\end{equation}
where the maximum is taken over all possible pure state decompositions $\{p_i,~|\psi_i\rangle_{AB}\}$ of $\rho_{AB}$.

For an $N$-qubit state $\rho_{AB_1\cdots B_{N-1}}\in H_A\otimes H_{B_1}\otimes\cdots\otimes H_{B_{N-1}}$, we denote $N_c(\rho_{A|B_1\cdots B_{N-1}})$ the negativity of the state $\rho_{A|B_1\cdots B_{N-1}}$, viewed as a bipartite state under the partition $A$ and $B_1,B_2,\cdots, B_{N-1}$. If
$N_c(\rho_{AB_i})\geq N_c(\rho_{A|B_{i+1}\cdots B_{N-1}})$ for $i=1, 2, \cdots, m$, and
$N_c(\rho_{AB_j})\leq N_c(\rho_{A|B_{j+1}\cdots B_{N-1}})$ for $j=m+1,\cdots,N-2$,
$\forall$ $1\leq m\leq N-3$, $N\geq 4$, then \cite{jzx1}
\begin{eqnarray}\label{la3}
&&N^x_c(\rho_{A|B_1B_2\cdots B_{N-1}})\nonumber\\
 &&\geq N^x_c(\rho_{AB_1})+\frac{x}{2} N^x_c(\rho_{AB_2})+\cdots+\left(\frac{x}{2}\right)^{m-1}N^x_c(\rho_{AB_m})\nonumber\\
 &&+\left(\frac{x}{2}\right)^{m+1}(N^x_c(\rho_{AB_{m+1}})
 +\cdots+N^x_c(\rho_{AB_{N-2}}))\nonumber\\
 &&+\left(\frac{x}{2}\right)^{m}N^x_c(\rho_{AB_{N-1}}),
\end{eqnarray}
for all $x\geq 2$. The inequality (\ref{la3}) is further improved that for $x\geq2$ \cite{jll}
\begin{eqnarray}\label{ne}
&&{N^x_c}(\rho_{A|B_1B_2\cdots B_{N-1}})\nonumber\\
 &&\geq {N^x_c}(\rho_{AB_1})+h {N^x_c}(\rho_{AB_2})+\cdots+h^{m-1}{N^x_c}(\rho_{AB_m})\nonumber\\
 &&+h^{m+1}({N^x_c}(\rho_{AB_{m+1}})
 +\cdots+{N^x_c}(\rho_{AB_{N-2}}))\nonumber\\
 &&+h^{m}{N^x_c}(\rho_{AB_{N-1}})
\end{eqnarray}
where $h=2^\frac{x}{2}-1$.

The negativity of assistance does not satisfy a monogamy relation in general. However, for an $N$-qubit generlized $W$-class state $|\psi\rangle_{AB_1\cdots B_{N-1}}\in H_A\otimes H_{B_1}\otimes\cdots\otimes H_{B_{N-1}}$, if $N_c(\rho_{AB_i})\geq N_c(\rho_{A|B_{i+1}\cdots B_{N-1}})$ for $i=1, 2, \cdots, m$, and
$N_c(\rho_{AB_j})\leq N_c(\rho_{A|B_{j+1}\cdots B_{N-1}})$ for $j=m+1,\cdots,N-2$,
$\forall$ $1\leq m\leq N-3$, $N\geq 4$, then the negativity of assistance $N_a(|\psi\rangle_{A|B_1\cdots B_{N-1}})$ of the state $|\psi\rangle_{AB_1\cdots B_{N-1}}$ satisfies the inequality \cite{jzx1},
\begin{eqnarray}\label{la4}
&&N_a^x(|\psi\rangle_{A|B_1B_2\cdots B_{N-1}})\geq \nonumber\\
&&N_a^x(\rho_{AB_1})+\frac{x}{2} N_a^x(\rho_{AB_2})+\cdots+\left(\frac{x}{2}\right)^{m-1}N_a^x(\rho_{AB_m})\nonumber\\
&&+\left(\frac{x}{2}\right)^{m+1}\left(N_a^x(\rho_{AB_{m+1}})+\cdots+N_a^x(\rho_{AB_{N-2}})\right)\nonumber\\
&&+\left(\frac{x}{2}\right)^{m}N_a^x(\rho_{AB_{N-1}}),
\end{eqnarray}
for all $x\geq2$.

In fact, to have a better characterization of the entanglement distribution among the subsystems, the monogamy inequalities satisfied by the negativity of assistance can be further refined and become tighter.
Taking into account the fact that for $N$-qubit generlized $W$-class states (\ref{gw}) \cite{jzx1},
\begin{eqnarray}\label{la5}
N_c(\rho_{AB_i})=N_a(\rho_{AB_i}),~~i=1,2,\cdots,N-1
\end{eqnarray}
where $\rho_{AB_i}=\mathrm{Tr}_{B_1\cdots B_{i-1}B_{i+1}\cdots B_{N-1}}(|\psi\rangle_{AB_1\cdots B_{N-1}}\langle\psi|)$, we have the following relations:

{[\bf Theorem 3]}.
For the $N$-qubit generalized $W$-class states $|\psi\rangle_{AB_1\cdots B_{N-1}}\in H_A\otimes H_{B_1}\otimes\cdots\otimes H_{B_{N-1}}$, if $N_c(\rho_{AB_{j_i}})\geq N_c(\rho_{AB_{j_{i+1}}\cdots B_{j_{m-1}}})$ for $i=1,2,\cdots,t$, and $N_c(\rho_{AB_{j_k}})\leq N_c(\rho_{AB_{j_{k+1}}\cdots B_{j_{m-1}}})$ for $k=t+1,\cdots,{m-2}$, $\forall$ $1\leq t\leq {m-3},~m\geq 4$, then the CRENOA satisfies
\begin{eqnarray}\label{th4}
&&N_a^x(\rho_{A|B_{j_1}\cdots B_{j_{m-1}}})\geq N_a^x(\rho_{AB_{j_1}}) \nonumber\\
&&+hN_a^x(\rho_{AB_{j_2}})+\cdots+h^{t-1}N_a^x(\rho_{AB_{j_t}}) \nonumber\\
&&+h^{t+1}\left(N_a^x(\rho_{AB_{j_{t+1}}})+\cdots+N_a^x(\rho_{AB_{j_{m-2}}})\right)\nonumber\\
&&+h^tN_a^x(\rho_{AB_{j_{m-1}}})
\end{eqnarray}
for all $x\geq2$, where $h=2^\frac{x}{2}-1$.

{\sf [Proof].}
For the $N$-qubit generalized $W$-class states $|\psi\rangle_{AB_1\cdots B_{N-1}}$, according to the definitions of $N_c(\rho)$ and $N_a(\rho)$, one has $N_a(\rho_{A|B_{j_1}\cdots B_{j_{m-1}}})\geq N_c(\rho_{A|B_{j_1}\cdots B_{j_{m-1}}})$. When $x\geq2$, we have
\begin{eqnarray}\label{pf4}
N_a^x(\rho_{A|B_{j_1}\cdots B_{j_{m-1}}})&&\geq N_c^x(\rho_{A|B_{j_1}\cdots B_{j_{m-1}}})
\geq N_c^x(\rho_{AB_{j_1}}) \nonumber\\
&&+hN_c^x(\rho_{AB_{j_2}})+\cdots+h^{t-1}N_c^x(\rho_{AB_{j_t}}) \nonumber\\
&&+h^{t+1}\left(N_c^x(\rho_{AB_{j_{t+1}}})+\cdots+N_c^x(\rho_{AB_{j_{m-2}}})\right)\nonumber\\
&&+h^t N_c^x(\rho_{AB_{j_{m-1}}})\nonumber\\
&&=N_a^x(\rho_{AB_{j_1}})
+hN_a^x(\rho_{AB_{j_2}})+\cdots+h^{t-1}N_a^x(\rho_{AB_{j_t}}) \nonumber\\
&&+h^{t+1}\left(N_a^x(\rho_{AB_{j_{t+1}}})+\cdots+N_a^x(\rho_{AB_{j_{m-2}}})\right)\nonumber\\
&&+h^tN_a^x(\rho_{AB_{j_{m-1}}}),
\end{eqnarray}
where we have used in the first inequality the relation $a^x\geq b^x$ for $a\geq b\geq 0,~x\geq 2$. Due to the (\ref{ne}), one gets the second inequality. The equality is due to relation (\ref{la5}).
\hfill \rule{1ex}{1ex}

As for $x\geq 2$, $h^t\geq (x/2)^t$ for all $1\leq t\leq m-3$, comparing with
the monogamy relations for CRENOA in (\ref{la4}), our formula (\ref{th4}) in Theorem 3 gives a tighter monogamy relation with larger lower bounds.
In Theorem 3 we have assumed that
some $N_c(\rho_{AB_{j_i}})\geq N_c(\rho_{AB_{j_{i+1}}\cdots B_{j_{m-1}}})$ and some
$N_c(\rho_{AB_{j_k}})\leq N_c(\rho_{AB_{j_{k+1}}\cdots B_{j_{m-1}}})$ for the $N$-qubit generalized $W$-class states.
If all $N_c(\rho_{AB_{j_i}})\geq N_c(\rho_{AB_{j_{i+1}}\cdots B_{j_{m-1}}})$ for $i=1, 2, \cdots, {m-2}$, then we have
the following conclusion:

{\bf [Theorem 4]}.
If $N_c(\rho_{AB_{j_i}})\geq N_c(\rho_{AB_{j_{i+1}}\cdots B_{j_{m-1}}})$ for $i=1, 2, \cdots, {m-2}$, we have
\begin{eqnarray}\label{th5}
N_a^x(\rho_{A|B_{j_1}\cdots B_{j_{m-1}}})\geq N_a^x(\rho_{AB_{j_1}}) +hN_a^x(\rho_{AB_{j_2}})+\cdots+h^{m-2}N_a^x(\rho_{AB_{j_{m-1}}})
\end{eqnarray}
for all $x\geq2$, where $h=2^\frac{x}{2}-1$.

{\it Example 2}. Let us consider the $N$-qubit pure $W$-class state,
\begin{eqnarray}\label{WN}
|W\rangle_{AB_1\cdots B_{N-1}}=\frac{1}{\sqrt{N}}(|10\cdots0\rangle+|01\cdots0\rangle+\cdots+|00\cdots1\rangle).
\end{eqnarray}
It is straightforword to check: $N_a^x(|W\rangle_{A|B_1\cdots B_{N-1}})=(\frac{2\sqrt{N-1}}{N})^x$, $N_a^x(\rho_{AB_1})=N_a^x(\rho_{AB_2})=\cdots=N_a^x(\rho_{AB_{N-1}})=(\frac{2}{N})^x$. Let us choose $N=5$. Then $N_a^x(|W\rangle_{A|B_1\cdots B_4})=(\frac{4}{5})^x$. From our result (\ref{th4}) we have $N_a^x(|W\rangle_{A|B_1\cdots B_4})\geq \left[3\cdot 2^\frac{x}{2}-2\right](\frac{1}{2})^x$, while from (\ref{la4}) one has $N_a^x(|W\rangle_{A|B_1\cdots B_4})\geq (\frac{3x}{2}+1)(\frac{1}{2})^x$, $x\geq2$. Obviously, our result is better than that in \cite{jzx1} with $x\geq2$, see Fig. 2.

\begin{figure}
  \centering
  \includegraphics[width=13cm]{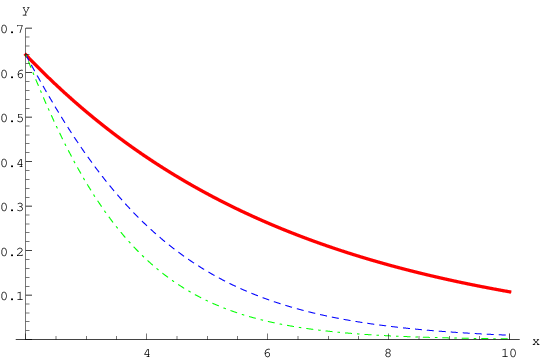}\\
  \caption{$y$ is the value of $N_a^x(|W\rangle_{A|B_1\cdots B_4})$. Solid (red) line is the exact value of $N_a^x(|W\rangle_{A|B_1\cdots B_4})$, dashed (blue) line is the lower bound of $N_a^x(|W\rangle_{A|B_1\cdots B_4})$ in (\ref{th4}), dot dashed (green) line is the lower bound in \cite{jzx1} for $x\geq2$.}\label{1}
\end{figure}

\section{conclusion}
Entanglement monogamy is a important property of multipartite systems. Tighter monogamy inequalities for $C_a^x(\rho_{A|B_{j_1}\cdots B_{j_{m-1}}})$,  $4\leq m \leq N$,  have been obtained for the $N$-qubit generalized $W$-class states with $x\geq 2$.
With a similar method, for $x\geq 2$, we also have investigated the $x$-power of negativity of assistance for the $N$-qubit generalized $W$-class states.
We have shown that these monogamy-type inequalities are tighter than the existing ones, thus give rise to the better restrictions of entanglement distribution among the qubits in generalized $W$-class states.
Nevertheless, for some classes of quantum states, similar to quantum entanglement, certain monogamy relations are also established.

\bigskip
\noindent{\bf Acknowledgments}\, \, This work is supported by the NSF of China under Grant No. 11675113.

\end{document}